\documentclass[prb,superscriptaddress,floatfix,a4paper,aps,twocolumn,nofootinbib]{revtex4-2}

\usepackage[T1]{fontenc} 

\usepackage{graphicx}
\usepackage{amsmath}
\usepackage{amssymb}
\usepackage{xcolor}

\usepackage{tabularx}
\usepackage{xspace}
\usepackage[hidelinks]{hyperref} 

\begin{document}

\author{Beilun Wu}
\affiliation{Laboratorio de Bajas Temperaturas y Altos Campos Magn\'eticos, Unidad Asociada UAM-CSIC, Departamento de Física de la Materia Condensada, Universidad Autónoma de Madrid, E-28049 Madrid, Spain}
\affiliation{Condensed Matter Physics Center (IFIMAC), Universidad Autónoma de Madrid, E-28049 Madrid, Spain}
\affiliation{Instituto Nicolás Cabrera (INC), Universidad Autónoma de Madrid, E-28049 Madrid, Spain}

\author{Andr\'es Mart\'inez}
\affiliation{Departamento de F\'\i sica and Instituto Universitario de Materiales de Alicante (IUMA), Universidad de Alicante, Campus de San Vicente del Raspeig, E-03690 Alicante, Spain.}

\author{Paula Obladen}
\affiliation{Laboratorio de Bajas Temperaturas y Altos Campos Magn\'eticos, Unidad Asociada UAM-CSIC, Departamento de Física de la Materia Condensada, Universidad Autónoma de Madrid, E-28049 Madrid, Spain}
\affiliation{Condensed Matter Physics Center (IFIMAC), Universidad Autónoma de Madrid, E-28049 Madrid, Spain}
\affiliation{Instituto Nicolás Cabrera (INC), Universidad Autónoma de Madrid, E-28049 Madrid, Spain}

\author{Marta Fern\'andez-Lomana}
\affiliation{Laboratorio de Bajas Temperaturas y Altos Campos Magn\'eticos, Unidad Asociada UAM-CSIC, Departamento de Física de la Materia Condensada, Universidad Autónoma de Madrid, E-28049 Madrid, Spain}
\affiliation{Condensed Matter Physics Center (IFIMAC), Universidad Autónoma de Madrid, E-28049 Madrid, Spain}
\affiliation{Instituto Nicolás Cabrera (INC), Universidad Autónoma de Madrid, E-28049 Madrid, Spain}

\author{Edwin Herrera}
\affiliation{Laboratorio de Bajas Temperaturas y Altos Campos Magn\'eticos, Unidad Asociada UAM-CSIC, Departamento de Física de la Materia Condensada, Universidad Autónoma de Madrid, E-28049 Madrid, Spain}
\affiliation{Condensed Matter Physics Center (IFIMAC), Universidad Autónoma de Madrid, E-28049 Madrid, Spain}
\affiliation{Instituto Nicolás Cabrera (INC), Universidad Autónoma de Madrid, E-28049 Madrid, Spain}

\author{Carlos Sabater}
\affiliation{Departamento de F\'\i sica and Instituto Universitario de Materiales de Alicante (IUMA), Universidad de Alicante, Campus de San Vicente del Raspeig, E-03690 Alicante, Spain.}

\author{Juan Jos\'e Palacios}
\affiliation{Condensed Matter Physics Center (IFIMAC), Universidad Autónoma de Madrid, E-28049 Madrid, Spain}
\affiliation{Instituto Nicolás Cabrera (INC), Universidad Autónoma de Madrid, E-28049 Madrid, Spain}
\affiliation{Departamento de Física de la Materia Condensada, Universidad Autónoma de Madrid, E-28049 Madrid, Spain}

\author{Isabel Guillam\'on}
\affiliation{Laboratorio de Bajas Temperaturas y Altos Campos Magn\'eticos, Unidad Asociada UAM-CSIC, Departamento de Física de la Materia Condensada, Universidad Autónoma de Madrid, E-28049 Madrid, Spain}
\affiliation{Condensed Matter Physics Center (IFIMAC), Universidad Autónoma de Madrid, E-28049 Madrid, Spain}
\affiliation{Instituto Nicolás Cabrera (INC), Universidad Autónoma de Madrid, E-28049 Madrid, Spain}

\author{Hermann Suderow}
\affiliation{Laboratorio de Bajas Temperaturas y Altos Campos Magn\'eticos, Unidad Asociada UAM-CSIC, Departamento de Física de la Materia Condensada, Universidad Autónoma de Madrid, E-28049 Madrid, Spain}
\affiliation{Condensed Matter Physics Center (IFIMAC), Universidad Autónoma de Madrid, E-28049 Madrid, Spain}
\affiliation{Instituto Nicolás Cabrera (INC), Universidad Autónoma de Madrid, E-28049 Madrid, Spain}

\title{Conductance of atomic size contacts of Ag and Au at high magnetic fields} 

\begin{abstract}
Electronic conduction at the atomic scale can be described by Landauer's formalism. In single atom point contacts of noble metals like Au and Ag, there is just one channel open between both electrodes and the conductance is very close to the quantum of conductance $G \approx G_0=\frac{2e^2}{h}$, with the factor of two coming from spin degeneracy. The magnetoconductivity of atomic size contacts has been studied for numerous systems, unveiling local Kondo screening, magnetic order and spin-polarized currents. However, these have been mostly performed in elements with multiple open conduction channels where $G$ differs from $G_0$. The realization of a magnetically active conductor with a single open channel remains difficult to achieve. Here we present measurements of the electronic conductance of single channel Au and Ag atomic‐size contacts in magnetic fields up to 20 Tesla. We observe a decrease in $G$ which goes up to about 15\% in many Au contacts at 20 T.  We perform calculations and find that pure Ag and Au do not present a strong field dependence of $G$, in agreement with previous results at smaller magnetic fields. We also find, however, that residual O$_2$ molecules attached close to the contact produce an an induced spin-polarized current, which leads to a decrease in $G$. We discuss the role of the magnetic response of the electrodes in the jump to contact. Our results suggest that single channel atomic size conductors with a sizeable response to a magnetic field can be built by combining noble metals and magnetically active molecular systems.
\end{abstract}

\maketitle

\section{Introduction}

Atomic-size contacts created between two sharp electrodes provide insight into atomic bonding and electronic conduction properties at the nanometric scale\,\cite{AGRAIT2003,bookCuevasScheer,Requist2016,Heinrich2021,Krans1995,Ohnishi1998,Yanson1998,PhysRevB.93.085437}. At this scale, the conductance is described by Landauer's formula, $G=G_0\sum_{i=1}^{N} T_i$, where $G_0=\frac{2e^2}{h}$ is the quantum of conductance, $i$ denotes the conduction channel number, $N$ is the number of opened channels and $0\leq T_i\leq 1$ the transparency of each channel\,\cite{Landauer1957,Nazarov}.  $N$ is related to the valence of the contacting atom\,\cite{Scheer1998}, and the coefficients $T_i$ serve as a "pin code" characterising the properties of the nanoscale conductor\,\cite{Scheer1998}. For instance, atomic-size contacts of monovalent Au and Ag exhibit a conductance plateau at $1G \simeq G_0$ ($N=1$ and $T_1\approx 1$) which is nearly independent of the geometry and shape of the nanocontact\,\cite{Agrait1993,Scheer1998,PhysRevB.66.085418,Cuevas1998b,PhysRevLett.87.026101,PhysRevB.88.161404,Cuevas1998a}. The influence of a magnetic field on the conductance of atomic-size contacts has been considered in Refs.\,\cite{Smit2001,Strigle2015,Rodrigues2003Co,Untiedt2004,Fernandez2005,Kinikar2017,PhysRevLett.115.036601,Brun2014,PhysRevB.70.064410,PhysRevB.71.220403,SUDEROW2003264,Doudin2008,Calvo2009,Strigle2015,Doudin2008}.

In nanocontacts of magnetic materials such as Fe and Co, spin-dependent transport due to ferromagnetic interactions results in strong variations of the conductance $G$ under applied magnetic fields\,\cite{Doudin2008,Calvo2009,Strigle2015,PhysRevB.88.161404,Vardimon2015,Rakhmilevitch2016,PhysRevB.88.245431,PhysRevB.93.085439,Chakrabarti2022,Hayakawa2016,PhysRevB.81.134402,PhysRevB.94.144431,PhysRevB.100.214439}. However, most atomic-size contacts that exhibit features associated with the application of a magnetic field are composed by elements with a valence different from one. Consequently, these contacts possess several conduction channels, $N\ge 1$, each being  partially open, $0< T_i< 1$. The "pin code" and overall conductance through these atomic-size contacts are highly sensitive to the atomic arrangement close to the contact\,\cite{Cuevas1998a,Cuevas1998b}. Here we measure the magnetic field dependence of the conductance of monovalent Au and Ag atomic-size contacts up to 20\,T. We also perform calculations showing that O$_2$ molecules, which exhibit high spin-orbit coupling, located sufficiently close to the contact yield spin-dependent conduction under magnetic fields (Fig.\,1(a)). Furthermore, we also discuss the influence of the magnetic field in the contact formation (Fig.\,1(b)).

\begin{figure*}[htb]
		\centering
		\begin{center}
			\includegraphics[width = 0.9\textwidth]{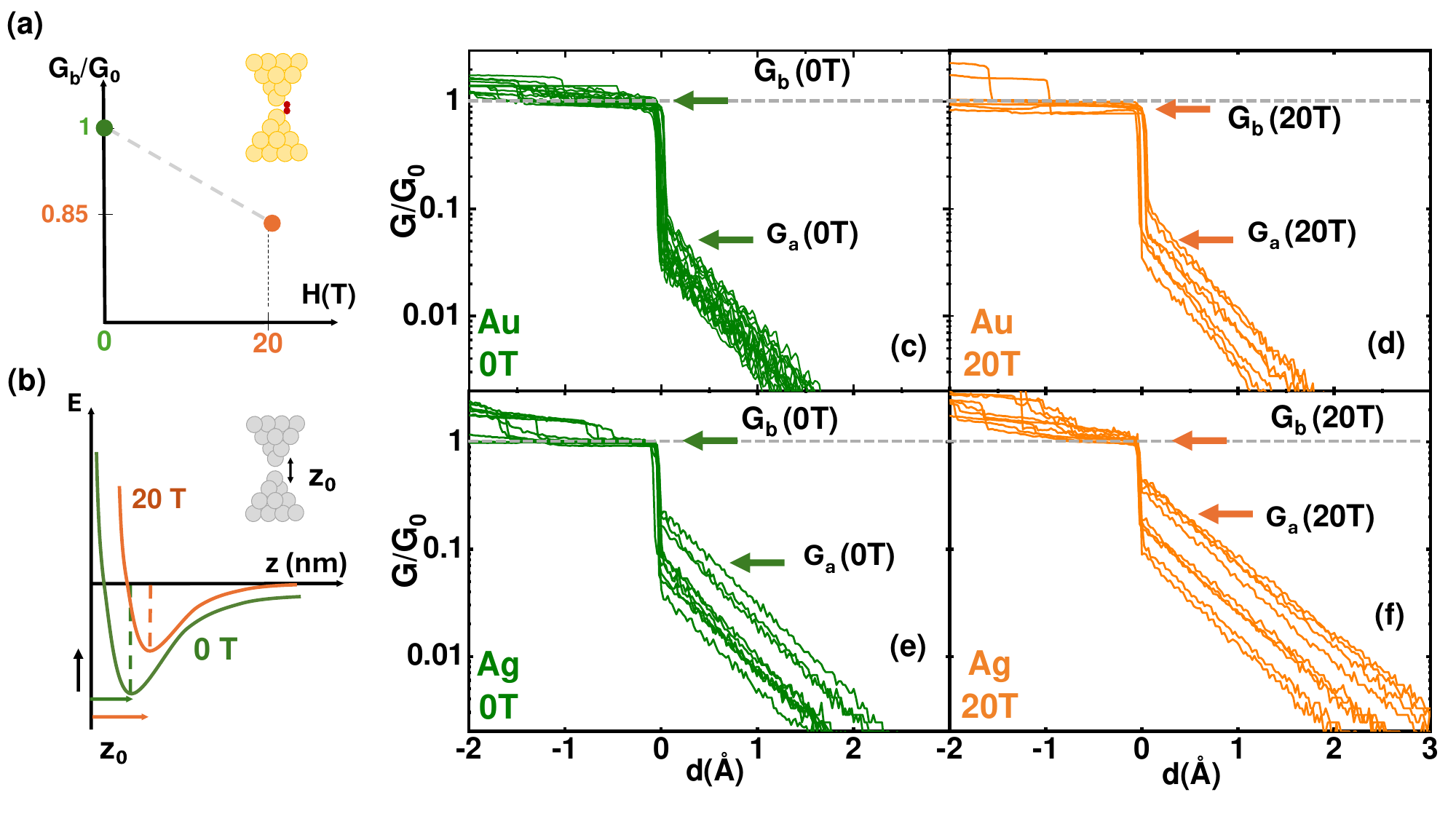}
		\end{center}
		\caption{(a) Schematic representation of the variation of the single-atom point contact conductance $G_b$ normalized to the quantum of conductance $G_0$ as a function of the magnetic field in Au (dashed line and green and orange points). We show in the upper right inset a molecular junction consisting of O$_2$ (red) attached to atomic Au electrodes (depicted in yellow). (b) Schematic representation of the binding energy as a function of the distance $z$ in Ag (green and orange lines) for zero magnetic field and for 20 T. The grey disks in the upper right inset represent Ag atoms, and the equilibrium distance $z_0$ is indicated by arrows. (c-f) We show as colored lines some representative measurements of the conductance $G$ (normalized to the quantum of conductance $G_0$) versus distance $d$ in Ag and Au. Here $d=0$ is defined as the point of contact formation; for $d>0$, the characteristic exponential dependence of the tunneling regime is observed. Curves are obtained by moving the z-position of the tip from the tunneling into the contact regime (from right to left in the figure). The arrows indicate the final conductance value in the tunneling regime, $G_a$, and the conductance of the single-atom contact, $G_b$. Data are presented for zero magnetic field (panels (a) and (c)) and under a magnetic field of 20\,T (panels (b) and(d)).}\label{FigureCurvesRepresentative}
		\end{figure*}
		
\section{Methods}

\subsection{Experimental}

We employed a cryogenic Scanning Tunneling microscope (STM) integrated with a 20 Tesla fully superconducting magnet from Oxford Instruments\,\cite{Marta_RSI}. The details of the STM hardware and software are provided in Ref.\,\cite{Fran_RSI}. To achieve clean and reproducible nanocontacts, the tip and sample were repeatedly indented at 4.2\,K. This process mechanically anneals the contact region, as described in Refs.\,\cite{PhysRevLett.108.205502,doi:10.1063/1.3567008,PhysRevLett.88.216803,PhysRevB.62.9962,PhysRevLett.93.116803} and facilitates the formation of well defined atomic-size contacts. Repeated indentation resets completely the contact area through the formation of different connecting necks at each contact and does not allow to follow a single contact as a function of the magnetic field, as made in Refs.\,\cite{Strigle2015,Rakhmilevitch2016,Hayakawa2016,PhysRevB.81.134402,PhysRevB.94.144431,PhysRevB.100.214439}. However, it provides access to atomic size contacts connnected to the bulk through necks of many different shapes. This approach also eliminates magnetostriction effects on the junction, as the experiment is reset at each magnetic field. Furthermore, by implementing the in-situ positioning device described in Ref.\,\cite{doi:10.1063/1.3567008} multiple locations were probed. We used a $10^6$ current-to-voltage amplifier with a tip-sample bias voltage of 50\,mV. All measurements were performed under a constant magnetic field, with the superconducting coil in persistent mode\,\cite{Marta_RSI}. The Au and Ag tips and samples were prepared from wires with a purity of 99.99\%. Inductively coupled plasma mass spectrometry revealed the following amounts of impurities: Si (24 ppm), Fe (17 ppm), Cu (5 ppm), Zn (33 ppm), Ag (100 ppm), Pd (5 ppm), Pt (12 ppm) and Co, Ni $<$ 1 ppm in our Au sample; in Ag, we find Au (20 ppm), Hg (6 ppm), Pb (2 ppm), Se (19 ppm), Zn (183 ppm), Cu (31 ppm), Fe (25 ppm), Al (51 ppm), along with traces of alkali elements. These impurity concentrations are sufficiently low to be considered negligible for the purposes of this study. At each magnetic field, tens of thousands of conductance versus distance curves were acquired, from which the conductance histograms presented below were constructed. In Fig.\,\ref{FigureCurvesRepresentative}(c-f), zero distance ($d=0$) is defined as the jump to contact, identified by an increase in conductance of more than 0.3\,$G_0$ within one picometer.

\subsection{Modelling the atomic structure of nanocontacts}
We generate the Au nanocontact structures discussed below through classical molecular dynamics (CMD) simulations of the pull-push process on a nanowire oriented along the (001) crystallographic direction, initially featuring a narrowed middle section. The simulations were performed with the Large-Scale Atomic/Molecular Massively Parallel Simulator (LAMMPS) \cite{LAMMPS} employing an embedded-atom-model (EAM) potential \cite{EAM_1, EAM_2}. In our pull-push process, opposite velocities of $\pm 0.04$  \AA/ps were applied to the upper and lower layers of the electrode, with steps of 1 ps. The pulling process was conducted over 250 ns and the pushing process over 254 ns, ensuring the formation and rupture of the junction in the majority of the 30 pull-push cycles simulated. The system's temperature  was mantained at $4.2$ $K$ in an NVT canonical ensemble with a Nosé-Hoover thermostat.

From a snapshot of the CMD simulation in which a dimer configuration was observed, the central 76 atoms were extracted. Here, a dimer is defined as a contact consisting of two pyramid-like electrodes connected by two atoms, wheras a monomer is characterized by a single atom contact. To prepare the structure for the conductance calculations (discussed below and shown in Fig.\,\ref{Figureteo}), the atoms in the outermost layers were repositioned to form an ideal (001) lattice. The remaining atoms, onto which an O$_2$ molecule was randomly added, were relaxed using density functional theory (DFT) as implemented  in Gaussian16 \cite{g16} within the unrestricted Local Spin Density Approximation (LSDA). For the central Au and oxygen atoms, a LANL2DZ basis set \cite{Lanl2dz_1, Lanl2dz_2} was employed, while the outermost layers of Au were modeled using the minimal CRENBS basis set \cite{Crenbs}  according to the crystalline structure implemented. 

\subsection{Universal binding curve for Au and Ag nanocontacts}

To calculate the interaction energy between two atoms bounded to electrodes, we constructed tetrahedral-shaped clusters resembling the nanocontact for Au and Ag (more detailes in Appendix A, Fig.\,\ref{FigSupp_ForceCurves}). The energy  $E$ as a function of the inter-cluster distance $z$ was determined using density functional theory (DFT) at the generalized gradient approximation (GGA) level with the Perdew-Burke-Ernzerhof (PBE)\,\cite{Perdew_Burke_Ernzerhof_1996} functional and the D3 version of Grimme’s dispersion with Becke-Johnson damping (GD3BJ)\,\cite{GD3BJ} implementation in Gaussian16\,\cite{g16}. An all-electron basis set, x2c-TZVPall\,\cite{xc2TZVPall}, was used for both Au and Ag, and scalar relativistic effects were introduced via second-order Douglas-Kroll-Hess (DKH) scalar relativistic calculation\,\cite{Douglas_Kroll_1974, Hess_1985, Hess_1986, Jansen_Hess_1989}. To perform the calculation, we first optimized the edge length of the regular cluster and then increased the distance between the tips of the two relaxed clusters.

\subsection{Transport caculations}

Non-Equilibrium Green's Functions (NEGF) calculations were performed to estimate the conductance across the junctions decribed above. These calculations were run using our ANT.Gaussian code\,\cite{ANT_1, ANT_2, ANT_3, PhysRevB.71.220403, ANT_code}. In this work, we employed a new implementation that enables the inclusion of a magnetic field in the z-direction in a self-consistent manner. 

\begin{figure*}[t]
		\centering
		\begin{center}
			\includegraphics[width = 0.95\textwidth]{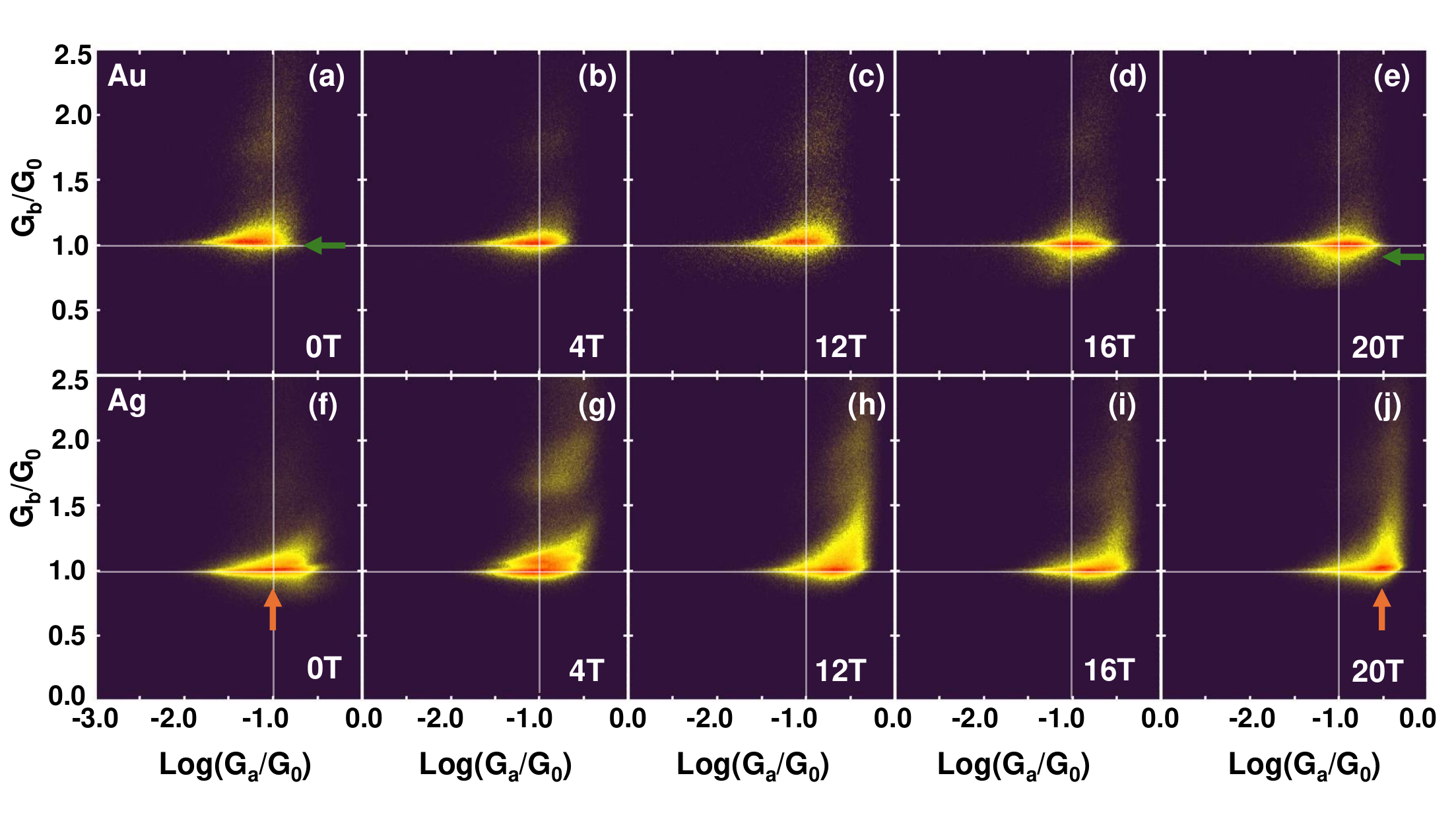}
		\end{center}
		\caption{We show two-dimensional histograms of $G_b/G_0$ vs $log(G_a/G_0)$ for various values of the magnetic field, indicated in each panel. The upper row displays the results for Au (a–e) and the bottom row (f–j) the results for Ag. The color scale shows occurrence of a certain conductance value from red (high) to blue (none) (cuts of the histogram and more details provided in Appendix D). Vertical and horizontal grey lines are visual guides. For the conductance of the single-atom point contact $G_b$ we mostly find $G_b=G_0$. In Au we observe that $G_b$ decreases below $G_0$ under applied magnetic fields, as highlighted by the green arrow. In Ag the conductance value immediately prior to the jump-to-contact event $G_a$ increases with the magnetic field, as highlighted by the orange arrow.}\label{FigureCurvesHisto}
		\end{figure*}
		
\section{Results}

In Figs.\,\ref{FigureCurvesRepresentative}(c-f) we present representative conductance versus distance curves for Au and Ag few-atom point contacts at 0 T and 20 T. We define $G_a$ as the conductance immediately prior to the jump-to-contact (i.e., the last conductance value recorded in the tunneling regime), and $G_b$ as the conductance value at the first contact (i.e.\,, the first value recorded immediately after the tunneling regime, as indicated by the arrows in Fig.\,\ref{FigureCurvesRepresentative}(c-f))\,\cite{Sabater2013,PhysRevB.93.085437,Sabater2018,Calvo2018,PhysRevB.106.125418,Dednam_2015}. In Au, $G_b$ decreases with the application of a magnetic field. In Ag, we observe a similar phenomenon, but less pronounced. In addition $G_a$ increases with the magnetic field. We can see the same effects by looking on large amounts of data from multiple contacts, shown in histograms in Fig.\,\ref{FigureCurvesHisto}. The histograms exhibit a weak tail for $G_b$ values below $G_0$, which is best visible at high magnetic fields and in Au (as indicated by the green arrow in Fig.\,\ref{FigureCurvesHisto}(e), see also more data in the Appendix D, Fig.\,\ref{FighistoGafixed}). We also observe that the maximum conductance value before contact, $G_{a,max}$ increases with the magnetic field, particularly for Ag (orange arrow in Fig.\,\ref{FigureCurvesHisto}(j)).

In Fig.\,\ref{FigureHistoGaGb}(a) we show the magnetic field dependence of the histogram of $G_a$ for contacts with $G_b=G_0$. We find an enhancement of $G_a$ with increasing magnetic field particularly for Ag. In Fig.\,\ref{FigureHistoGaGb}(b) we plot the fraction of atomic-size contacts with $G_b$ below $0.85G_0$ relative to the total number of contacts at each magnetic field. In Au, the proportion of contacts exhibiting conductance values that are 15\% lower than $G_0$ increases with the magnetic field.

\subsection{Magnetoresistance in Au nanocontacts.}

Thus, particularly for Au, our results show that the conductance $G_b$ for single-atom contacts is reduced below $G_0$ under high magnetic fields, while $G_a$ remains essentially invariant with respect to magnetic field.

The value of $G_b$ is the only parameter that produces a plateau in the conductance versus distance curves (Figs.\,\ref{FigureCurvesRepresentative}(c-f)), resulting in a well defined peak at $G_b\approx G_0$ in the histograms (Fig.\,\ref{FigureCurvesHisto}). Sometimes, multi-atom contacts are formed in the jump to contact process. These produce histogram counts at $G_b$ well above $G_0$\,\cite{Calvo2018,Sabater2018}. However, the contributions to the histogram for $G_b\leq G_0$ are nearly absent (see green arrow in Fig.\,\ref{FigureCurvesHisto}(a)). The abrupt nature of the contact formation causes a sudden increase in conductance, as shown in Figs.\,\ref{FigureCurvesRepresentative}(c-f). At high magnetic fields (green arrow in Fig.\,\ref{FigureCurvesHisto}(e)), we observe that the histogram extends into the region where $G_b < G_0$.

\subsection{Magnetic field induced modifications in bonding of Ag nanocontacts.}

At zero magnetic field, it was found previously that $G_a$ is generally larger in Ag than in Au\,\cite{Calvo2018,PhysRevLett.47.675,PhysRevLett.87.266101,Rodrigues2000Au}. As we can see in Figs.\,\ref{FigureCurvesRepresentative}(c,e), when the two electrodes approach each other, the conductance initially increases exponentially with decreasing distance. The interplay between the elastic interaction of the contacting atoms with the atom at the apex and atomic binding between the two atoms at both apexes leads to an abrupt jump into contact. As $G_b\approx G_0$, $G_a$ quantifies the separation between the electrodes where the contact forms. At zero field, $G_a$ is significantly smaller in Au than in Ag, because the formation of a two-atom bond between the atoms at the apexes occurs at larger separation distances in Au than in Ag\,\cite{Calvo2018,PhysRevLett.47.675,PhysRevLett.87.266101,Rodrigues2000Au}. This is due to the larger binding energy in Au, arising from differences in the interatomic potential (discussed with more detail in Appendix A and in Refs.\,\cite{Sabater2013,PhysRevB.93.085437,Sabater2018,Calvo2018,PhysRevB.106.125418,Dednam_2015}). In the histogram shown in Figs.\,\ref{FigureCurvesHisto}(a,f) we see that the peak of the histogram on the x-axis occurs for a smaller $G_a$ value in Au than in Ag.

When we apply a magnetic field, we find that the peak in the histogram in $G_a$ occurs generally for higher values of $G_a$. Particularly for Ag, there is a clear shift at the highest magnetic fields (orange arrow in Fig.\,\ref{FigureCurvesHisto}(j)).
 
The magnetic field induced changes in $G_b$ are weaker for single atom contacts of Ag than those for Au. In contrast, the conductance just before the jump to contact, $G_a$, increases more strongly with the magnetic field in Ag.

\begin{figure}[htbp]
		\centering
		\begin{center}
			\includegraphics[width = \columnwidth]{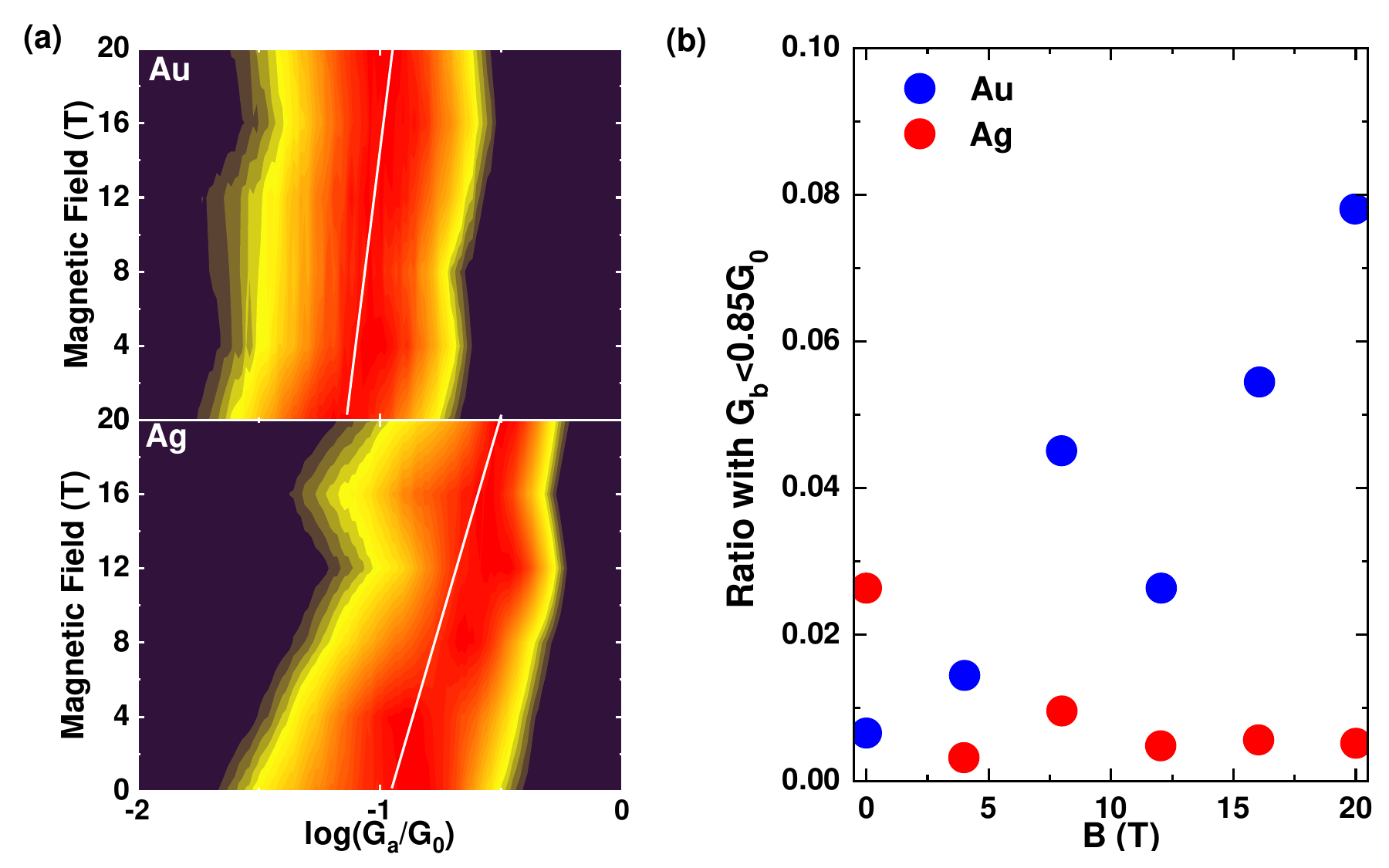}
		\end{center}
		\caption{(a) Color scale histograms of the conductance just before the jump-to-contact, $G_a$, for contacts with $G_b=1$, plotted as a function of the magnetic field for Au (top panel) and for Ag (bottom panel). White lines serve as visual guides. Red colour represents frequent occurence and the dark blue no occurrence. (b) The number of single-atom point contacts with $G_b<0.85 G_0$, as extracted from the histograms in Fig.\,\ref{FigureCurvesHisto}, is displayed as a function of the magnetic field (blue points for Au and red points for Ag).}\label{FigureHistoGaGb}
\end{figure}	

\section{Discussion}
\subsection{Spin dependent transport at high magnetic fields in atomic size contacts of gold}

Let us first consider the magnetic field-induced decrease of $G_b$ below $G_0$. As detailed in Appendix A, neither Au nor Ag exhibit intrinsic magnetic properties sufficiently strong to elicit a magnetic response in the conductance of atomic-sized contacts ($G_b$). In noble metal surfaces, there are Shockley-type surface states which are spin degenerate due to spin-orbit interaction and absence of inversion symmetry at the surface\,\cite{FriedrichReinert_2003,Reinert2004}. This requires a perfect atomically flat and ordered surface, often prepared in-situ by repeated sputtering and annealing.  The atomically flat surface required for such surface states is not present in the environment of our single atom point contacts. Furthermore, the spin splitting is only significant in presence of heavy atoms to enhance the spin orbit coupling\,\cite{PhysRevLett.98.186807,PhysRevB.72.045419}. In addition, previous calculations of the conductance of gold nanostructures show that spin related transport is not present in the geometries and bias voltage range we are considering in our work\,\cite{Dednam2023}. Thus, we can exclude spin phenomena in the conductance of these nanocontacts.

The experiments were conducted under cryogenic vacuum conditions, ensuring that no extraneous elements or large gas-phase molecules can contaminate the contact area. However, residual air (primarily N$_2$, O$_2$ and water) can condense on the sample surface, interacting with the contact. Since N$_2$ and water are non-magnetic, O$_2$ remains the only magnetically active molecule, which can be held responsible for the observed effects. As we discuss now, O$_2$ molecules could reach the contact region, thereby modifying the electronic conduction through the atomic size contact under high magnetic fields.

In Ag, O$_2$ is chemisorbed, which fixes the molecules on their adsorption place at the surface. Then, O$_2$ molecules might have a tendency to stick and the repeated indentation process does not scramble their positions. In contrast, O$_2$ is predominatly physisorbed on Au\,\cite{doi:10.1021/acs.chemrev.7b00217}. When an O$_2$ molecule approaches the contact region in Au, the reactivity increases due to the decreased coordination number of Au atoms at the contact\,\cite{PhysRevB.66.081405,doi:10.1021/ja312223t,PhysRevLett.96.026806,Thijssen_2008}. Under these conditions, the O$_2$ molecule has a stronger tendency to stick and could be incorporated into the contact, thereby contributing to the conduction behaviour in the single-atom point contact regime.

\begin{figure}[htbp]
		\centering
		\begin{center}
			\includegraphics[width = \columnwidth]{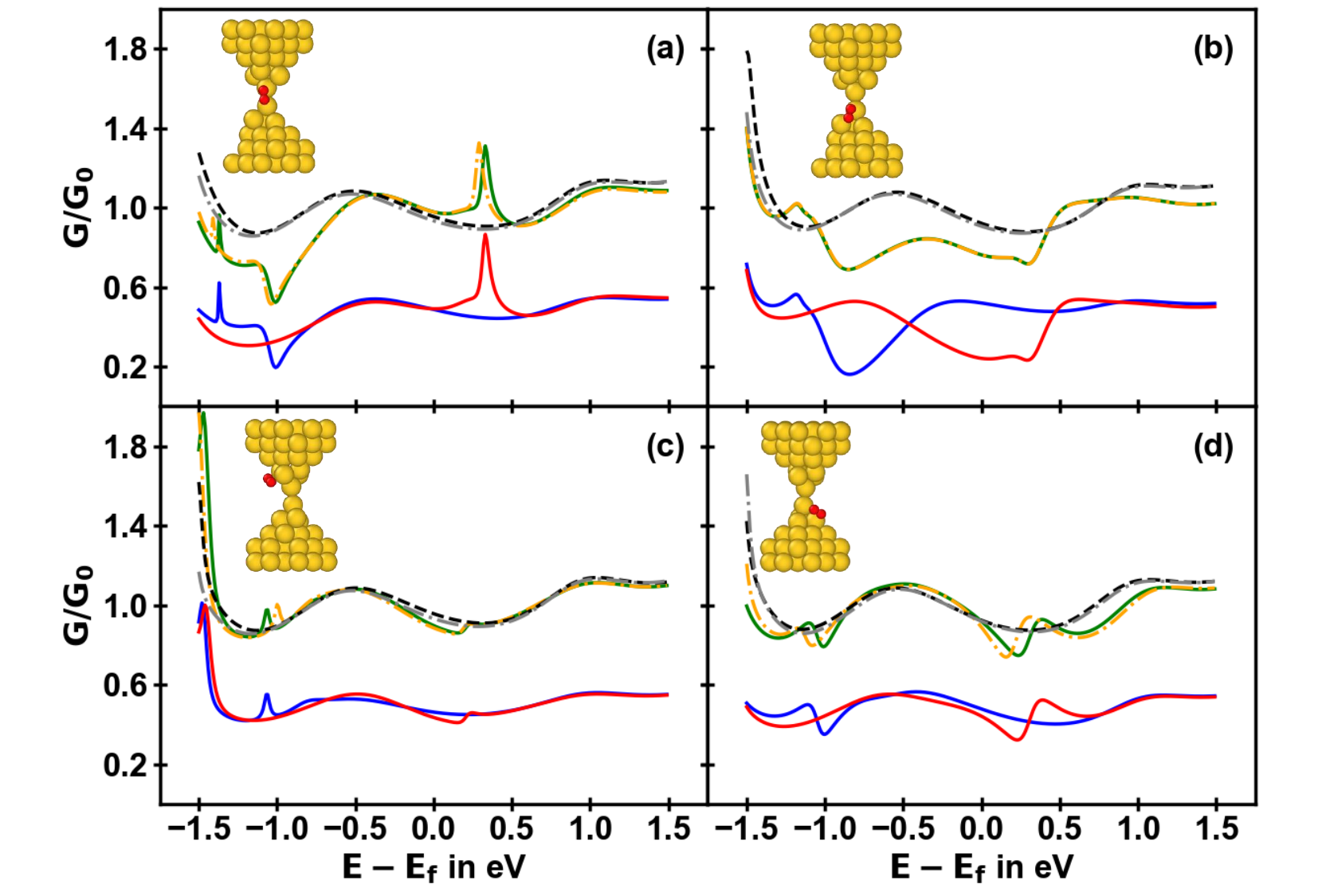}
		\end{center}
		\caption{(a–d) Calculated conductance $G$ normalized to the quantum of conductance $G_0$ as a function of energy (relative to the Fermi energy) for atomic-sized Au point contacts. The solid lines represent spin-resolved conductance: red for spin-up and blue for spin-down channels. The black dashed line corresponds to the conductance of a bare single-atom Au contact at zero magnetic field, while the grey dashed line shows the same configuration under an applied magnetic field of 20 T. The green and yellow dashed lines show the conductance for single-atom Au contacts with an O$_2$ molecule adsorbed near the contact region, at zero field and 20 T respectively. The specific atomic arrangements of the O$_2$ molecule (red disks) considered in each panel are shown in the top-left insets of (a–d). The presence and position of the O$_2$ molecule leads a notable spin-dependent conductance reduction near the Fermi level when the O$_2$ molecule is directly attached to the contacting atoms.}\label{Figureteo}
\end{figure}

We calculated the electronic conduction through atomic-size Au structures incorporating an O$_2$ molecule positioned at various locations near a single-atom point contact. We employed a molecular dynamics to model the formation of the nanocontacts. From this simulation, we selected a geometry exhibiting a single-atom contact. The single atom contact consists of a so-called dimer configuration in which two atoms join at the apex of a contacting structure, as shown in the insets of Fig.\,\ref{Figureteo}. We then added an O$_2$ molecule at different locations in its vicinity, and relaxed each configuration using density functional theory (DFT) (see insets of Fig.\,\ref{Figureteo}). Finally, we computed the conductance of the relaxed geometries as a function of the energy. These calculations were performed both in the absence of an external magnetic field and with a 20 T  applied magnetic field along the z-direction. The resulting conductance vs energy curves are presented in Fig.\,\ref{Figureteo}.

In atomic-size contacts of Au, prior to the incorporation of O$_2$, the conductance at the Fermi level ($E_F$) is approximately equal to $G_0$, with no significant dependence on the applied magnetic field, as shown by the black dashed lines in Fig.\,\ref{Figureteo}(a-d) for zero field and by the grey dashed lines for 20 T. When adding the O$_2$ molecule, different features appear in the conductance vs energy at aproximately 0.5\,eV above $E_F$ and 1\,eV below $E_F$. These features are due to the LUMO and HOMO levels of the O$_2$ molecule (peaks on the green and yellow dashed lines in Fig.\,\ref{Figureteo}(a-d)). These features are shifted in energy by a magnetic field. However, these shifts occur very far from the Fermi level $E_F$ and do not significantly influence the conductance measured in the experiment.

The conductance at the Fermi level $E_F$ is strongly sensitive to the position of the O$_2$ molecule. When O$_2$ is bonded directly to the two contacting atoms (Fig.\,\ref{Figureteo}(b)), we find a significant reduction of the conductance value below $G_0$. This decrease is attributed to a reduced spin-up current (red line in Fig.\,\ref{Figureteo}(b)), resulting in an overall conductance of $G\approx 0.8 G_0$. When located at this position, the O$_2$ molecule presents a magnetic moment of 1.74 $\mu_B$. Furthermore, $G$ displays a sizeable spin asymmetry. For other positions of the O$_2$ molecule, the conductance at the Fermi level remains close to $G\approx G_0$ and does not decrease much more than a few \%, evidencing that the molecule's effect on the electronic transport is highly dependent on its bonding configuration. The calculated spin magnetic moments are 1.21 $\mu_B$ in Fig.\,\ref{Figureteo}(c) and 1.3 $\mu_B$ Fig.\,\ref{Figureteo}(d) (see Appendix B for more details).

The decreased conductance of $G\approx 0.8 G_0$ observed in Fig.\,\ref{Figureteo}(b) aligns to the experimental findings: a reduction of $G_b$ by 15\% under 20 T. These calculations demonstrate that the presence of  O$_2$ can induce spin-dependent transport, leading to conductance values well below $G_0$ in certain atomic-size contacts. However, it is important to note that the green (zero field) and yellow (20 T) curves in Fig.\,\ref{Figureteo}(b) are remarkably similar, indicating that the conductance reduction itself is not directly altered by the magnetic field once the O$_2$ molecule is in place. This observation prompts the question of why contacts incorporating O$_2$ appear more frequently under high magnetic fields. The answer may lie in the role of magnetic field-induced spin alignment of O$_2$ adsorption process. Previous studies on reactive metal surfaces have shown that the orientation of the O$_2$ molecules relative to the surface significantly affects the overlap of molecular orbitals with the substrate\,\cite{KURAHASHI201629,10.1063/1.5111057,Luntz19884381,Zhao2023,PhysRevLett.68.624,PhysRevLett.94.036104,PhysRevB.55.15452} (see also Appendix B). Our experiment suggests that the magnetic field induces the adsorption, or "sticking" of the O$_2$ molecule, thereby increasing the probability of forming contacts where the O$_2$ molecule is directly incorporated in the junction.

To analyze this point, we estimated the magnetic anisotropy energy (MAE) of the O$_2$ molecule using a model described in Appendix C. We find that the spin orbit coupling (SOC) of Au is transferred to the O$_2$ molecule, resulting in a significant MAE of 6.5 $\times$ 10$^{-4}$\,eV when the molecule is aligned with the direction of the current flow (i.e.\, parallel to the contact axis). By contrast, the MAE is much smaller, of 8 $\times$ 10$^{-5}$\,eV, when the molecule is oriented perpendicular to the contact. This corresponds to a transverse easy axis of magnetization. This value of MAE, along with the favored orientation of the magnetic moment, changes with the position of the molecule. Since the direction of the applied magnetic field remains constant in the experiment, it produces a tendency to favor atomic arrangements where the magnetic moment of O$_2$ aligns with the magnetic field direction---most notably when O$_2$ is attached vertically between the two contacting atoms, as in Fig.\,\ref{Figureteo}(b). This eventually promotes atomic arrangements with partially spin polarized transport and a conductance of atomic-size contacts below $G_0$. 

\subsection{Contact formation at high magnetic fields}

In this section, we address the magnetic field-induced increase observed in $G_{a,max}$ which is most pronounced for Ag contacts. The conductance value $G_a$ depends on the conditions under which the two apex atoms form a bond. The interatomic bonding process can be described by the total energy $E(z)$ of the two-atom system as a function of their separation distance $z$, and we can write $E(z)=E_{UBC}(z)+E_{Elastic}(z)+E_{Magnetic}$. The position of the atoms at the apex is determined by the minimum in $E(z)$. $E_{UBC}(z)$ is the so-called universal binding curve, which describes the short-range attractive interaction between atoms, and is given by\,$E_{UBC}(z)=\alpha(z-z_0)e^{-\beta(z-z_0)}$\,\cite{PhysRevLett.47.675,PhysRevLett.87.266101,Rodrigues2000Au}. Here, $z_0$ is the equilibrium bonding distance in  the absence of external infuences, and $\alpha$ and $\beta$ are parameters determined from DFT calculations (see Appendix A). From these parameters we find that the binding energy ($E_0$), defined as the minimum of $E_{UBC}(z)$, is approximately 0.8 eV larger for Au than for Ag. Thus, the contributions form the elastic ($E_{Elastic}(z)$) and magnetic ($E_{Magnetic}$) terms, are more influential in Ag than in Au.

The elastic contribution, $E_{Elastic}(z)=\frac{k}{2}(D-z)^2$, accounts for the nanometric metallic constrictions holding the apex atoms \,\cite{PhysRevLett.100.175502}. Here, $k$ is the spring constant and $D$ is the externally controlled electrode separation that resembles the spring length (see Appendix A) \,\cite{PhysRevLett.100.175502,PhysRevLett.87.026101}. The observed variability in $G_a$ arises from differences in atomic configurations across contacts, leading to a distribution of values of $k$\,\cite{PhysRevLett.100.175502}.

At zero magnetic field, $G_{a,max}$ is higher for Ag than for Au. The minimum of $E(z)=E_{UBC}(z)+E_{Elastic}(z)$ occurs at larger atomic separations for Ag. This shift is also related to the lower binding energy in Ag compared to Au\,\cite{Calvo2018,PhysRevLett.108.205502}.

When we apply a magnetic field we observe an increase in $G_{a,max}$, particularly for Ag, exceeding the spread in $G_{a}$ for a fixed magnetic field. From the field-induced variation in $G_{a,max}$ we can estimate the decrease in electrode separation induced by the magnetic field and we find approximately 0.3\,\AA\,in Ag (using the conductance vs distance curves in the tunneling regime) under a field of 20\,T. The observed change in $G_{a,max}$ suggests that $E_{Magnetic}$ varies with the magnetic field. With $E_{Magnetic}$ of the order of hundreds of meV the bonding process is sizeably modified in Ag, while leaving Au mostly unaffected. The observed behaviour suggests an enhanced anisotropic magnetic susceptibility close to the contact region in nanosize constrictions. The anisotropy in the magnetization response for fields applied parallel versus perpendicular to the junction, denoted $\delta \mathbf{M}$, leads to a magnetic field-dependent torque $\boldsymbol{\tau} = \delta \mathbf{M} \times \mathbf{B}$\,\cite{PhysRevB.48.17001}, which produces a (rotational) magnetic force and contributes to $E_{Magnetic}$ as follows: $E_{Magnetic}\approx \chi_{anis}V\frac{B^2}{\mu_0} sin(\theta)$, with $\theta$ the angle between the magnetization and the magnetic field, $\chi_{anis}=\chi_{\vert\vert}-\chi_{\perp}$, where $\chi_{\vert\vert}$ and $\chi_{\perp}$ are the susceptibilities parallel and perpendicular to the magnetic field and $V$ the volume over which the torque is acting. Using bulk values (see Appendix A), we find $E_{Magnetic}$ of the order of $\mu$eV. It was shown in Ref.\,\cite{PhysRevLett.111.127202} that the anisotropy of the susceptibility $\chi_{anis}$ of small rods of Au is significantly enhanced with respect to the bulk value due to mesoscopic circular currents induced by the magnetic field. The enhancement was found to be of at least three order of magnitude, depending on the shape and size of the rods. Similar effects have been found in both Au and Ag nanoparticles\,\cite{doi:10.1021/nl073129g}. Furthermore currents through surface states on rod-like structures were shown to provide an anisotropic susceptibility about an order of magnitude larger than the bulk\,\cite{PhysRevLett.111.127202,doi:10.1021/nl073129g,Hernando_2014}. The shape of the contacting areas surrounding the atomic can be related to pyramidal or conic like atomic size arrangements\,\cite{PhysRevLett.81.4448,PhysRevB.56.2154,PhysRevB.61.12725}. Our results suggest that torques arising from anisotropic surface currents near the contact region contribute to $E_{Magnetic}$ and modify the zero field energy balance. These torques result in a magnetic field-induced decrease in the position of equilibrium for Ag, thereby increasing $G_{a,max}$.

We finally note that there are features in the conductance histogram (Fig.\,\ref{FigureCurvesHisto}) which depend on the magnetic field and remain difficult to address at present. There are sizeable modifications of the histograms for $G_b/G_0$ around 2. The number of atomic size contacts with values significantly above $G_0$ seems to be more pronounced in some runs. It is unclear if this can be related to a certain shape of the contacting electrodes, or to features which depend on the magnetic field. Generally, the spread of conductance values with $G_b/G_0$ well above one seems to increase with the magnetic field in Ag. Atomic size contacts with $G_b$ well above $G_0$ are formed by groups of atoms, instead of dimers\,\cite{Calvo2018,Sabater2018}. More detailed studies of the behavior of such multiple atom contacts under magnetic field could shed light in better understanding these features.

\section{Summary and conclusions}

In summary, we have studied the conductance properties of Au and Ag atomic-size contacts under magnetic fields up to 20 T. Our results show that high magnetic fields modify the conduction properties and the atomic environment near the contact region. We observe a conductance reduction under high magnetic fields, which goes up to approximately 15\% in Au contacts, and which we associate to the presence of O$_2$ molecules attached to the contacting atoms. We also observe a magnetic field induced modification of the jump to contact process which we attribute to a magnetic response of the nanosized areas close to the contact. Our findings show that the interplay between magnetic fields, surface chemistry, and electronic transport at the atomic scale in monovalent metals can lead to spin dependent transport at atomic scale.

\section*{Acknowledgements}
We acknowledge discussions with C. Untiedt and N. Agraït. This work was supported by the Spanish Research State Agency (PID2020-114071RB-I00, PID2023-150148OB-I00, PDC2021-121086-I00, TED2021-130546B\-I00, TED2021-131323B-I00, PID2022-141712NB-C21 and CEX2023001316-M), by the Comunidad de Madrid through program NANOMAGCOST-CM (Program No.S2018/NMT-4321),  Generalitat Valenciana through CIDEXG/2022/45 and PROMETEO/2021/017, the EU through grant agreement No 871106 and by the European Research Council PNICTEYES through grant agreements Pnicteyes 679080 and VectorFieldImaging 101069239. The QUASURF project (ref. SI4/PJI/2024-00199) funded by the Comunidad de Madrid through the agreement to promote and encourage research and technology transfer at the Universidad Autónoma de Madrid is also acknowledged. This work forms part of the Advanced Materials program and was supported by MCIN with funding from European Union NextGenerationEU (PRTR-C17.I1) and by Generalitat Valenciana (MFA/2022/045). We acknowledge collaborations through EU program Cost CA21144 (superqumap). We acknowledge SEGAINVEX for design and construction of cryogenic equipment and SIDI for support in sample characterization.

\paragraph*{Data availability.} The data that support the findings of this article are openly available \cite{DataPaper}.

\section*{Appendix A: Universal binding curve and other contributions to the energy}

\begin{figure}[htbp]
    \centering
    \includegraphics[width = 0.95\columnwidth]{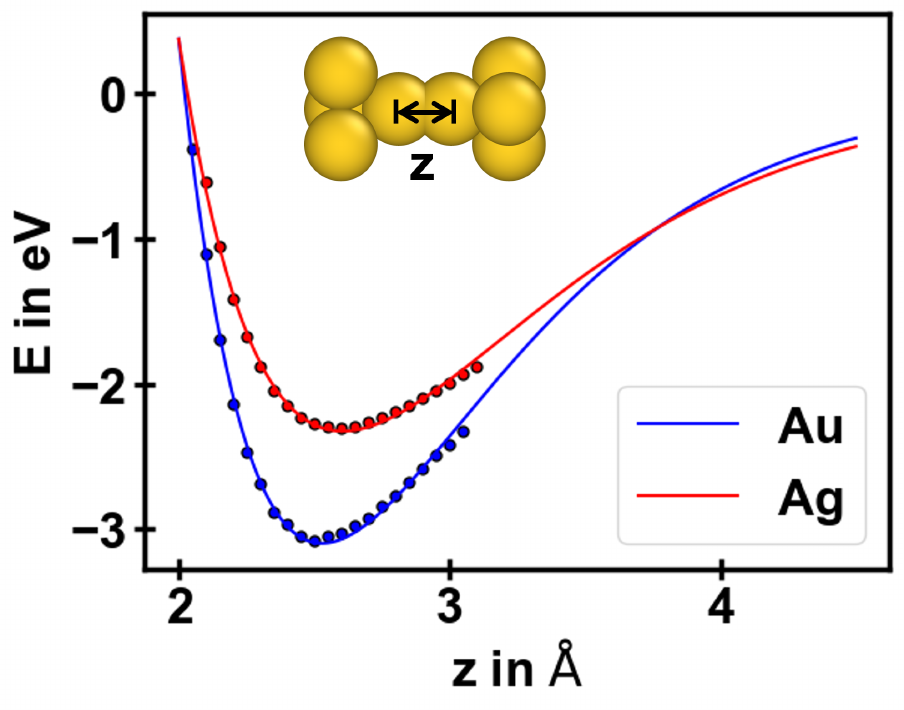}
    \caption{Binding energy as a function of the distance between the centers of the atoms located at the tip apex, obtained using the atomic arrangement shown in the inset. Filled circles are results of the calculation, and solid lines fits to the universal binding curve  (we used the parameters given in Table\,\ref{TabSupp_fit}).}
    \label{FigSupp_ForceCurves}
\end{figure}

We show in Fig.\,\ref{FigSupp_ForceCurves} the calculated binding energy for the nanocontact depicted in the inset as yellow disks. Results for Au and Ag are shown by red and blue filled circles, respectively. The overall shape of the energy versus distance curve is similar for both metals. The equilibrium bonding distances, i.e., the position of the energy minima ($z_0$ in Fig.\,\ref{FigureCurvesRepresentative}(b)), are comparable for both elements. However, the absolute value of the energy at the minimum is significantly smaller in Au than in Ag (see also Table\,\ref{TabSupp_fit}).

The points obtained by the calculation can be fitted using the universal binding curve\,\cite{PhysRevLett.47.675,PhysRevLett.87.266101,Rodrigues2000Au}. From the fitted parameters we extract the equilibrium bond distance $d=z_0+1/\beta$, the binding energy $E_0=-\alpha/\beta e$ and the breaking force $F_0=\alpha/e^2$ (or the slope at the inflection point of $E_{UBC}(z)$). These values are summarized in Table\,\ref{TabSupp_fit} and are consistent with pevious works\,\cite{PhysRevLett.100.175502,PhysRevLett.87.266101,Calvo2018,HAMMER200071}. Again, we note that the equilibrium bond distance $d$ is similar for both elements, but the breaking force $F_0$ and the binding energy at the minimum $E_0$ differ significantly, being substantially smaller for Ag than for Au.

\begin{figure}[htbp]
    \centering
    \includegraphics[width = 0.95\columnwidth]{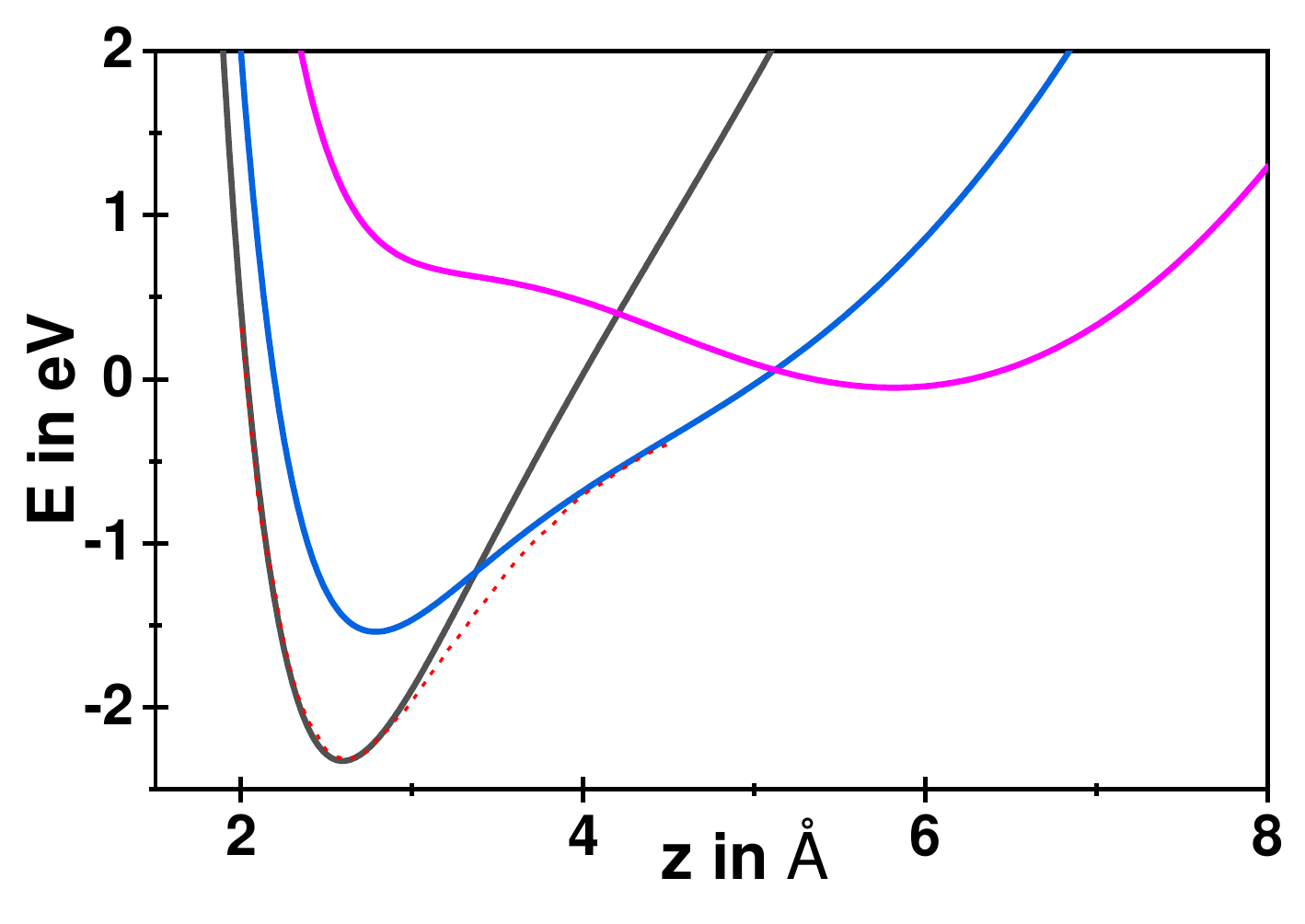}
    \caption{Energy $E_{UBC}+E_{Elastic}$ as a function of the atomic separation $z$, obtained for different values of the distance between electrodes $D$, for atomic contacts formed between Ag electrodes. The electrodes far apart in the magenta curve (weak interaction). The blue curve corresponds to electrodes approaching contact. The grey curve depicts a well-formed contact. The universal binding curve for Ag is shown by dashed red line.}
    \label{FigSupplModel}
\end{figure}

As discussed above, the total energy in the experimental setup must include the contribution from the metallic leads, which can be modeled by an elastic term $E_{Elastic}=\frac{k}{2}(D-z)^2$, with $k$ being a spring constant and $D$ the position of the nanoscale leads attached to the contact\,\cite{PhysRevLett.100.175502,PhysRevLett.87.026101}. Such an additional quadratic potential leads to a considerable modification of the energy landscape when added to to $E(z)$. This is shown in Fig.\,\ref{FigSupplModel}. The minimum of the total energy varies when modifying $D$. For large $D$, the elastic term dominates, and the energy minimum aligns with the minimum of $D$ (magenta curve in Fig.\,\ref{FigSupplModel}). When $D$ decreases, the contribution from the universal binding curve becomes increasingly relevant. At intermediate values of $D$, the combined energy profile can eventually exhibit two minima. This explains the hysteretic behavior of the contact formation, with one minimum for decreasing $D$ (contact formation) and another minimum for increasing $D$ (rupture of the contact). The jump to contact occurs at a critical value $D_0$, and the conductance immediately before this transition defines $G_{a,max}$ (blue and black curves in Fig.\,\ref{FigSupplModel})\,\cite{PhysRevLett.100.175502}.

\begin{table}[htbp]
\caption{Parameters of the universal binding curve obtained by fitting the model described in Fig.\,\ref{FigSupp_ForceCurves}.}
\label{TabSupp_fit}
\begin{tabular}{lccc} 

\hline \hline
\noalign{\vskip 1mm} 
Element & $d=z_0+1/\beta$ (\AA) & $E_0=-\alpha/\beta e$ (eV) & $F_0=\alpha/e^2$ (nN) \\ \hline
Au                          & 2.53 $\pm$ 0.02                 & -3.1 $\pm$ 0.2              & 2.24 $\pm$ 0.2 \\       
Ag                          & 2.6 $\pm$ 0.02               & -2.33 $\pm$ 0.2              & 1.49 $\pm$ 0.2    \\  \hline
\end{tabular}
\end{table}

When considering many different contacts, values for $D$ and $k$ are similar for Au and Ag. However, $G_{a,max}$ differs notably between the two metals at zero magnetic field. As shown in the conductance histograms in the left panels of Fig.\,\ref{FigureCurvesHisto} in the main text, the value measured for $G_{a,max}$ is higher for Ag than for Au. This observation is consistent with Ref.\,\cite{Calvo2018} and can be primarily attributed to the difference in the binding energy $E_0$ for the two elements (Fig.\,\ref{FigSupp_ForceCurves}). A relevant consequence for the present study is that the mechanical response of nanocontacts is more strongly influenced by the elastic term in Ag than in Au. This is because the absolute value of the binding energy $|E_0|$ is approximately 0.8 eV lower in Ag, so that the elastic contribution is more relevant in the contact formation for Ag than for Au.

We note that the spring constant $k$ is related to the elastic constants via $k\approx \frac{E_{Y} z_0}{1-\nu^2}$, where $E_{Y}$ is the Young's modulus and $\nu$ the Poisson's ratio\,\cite{PhysRevLett.100.175502}. The Young's modulus is given by the elastic constant $E_{Y}=C_{11}$.  There are sizeable differences in $C_{11}$ and $C_{12}$ among Ag and Au\,\cite{PhysRev.111.707,Qu1m}. These differences affect derived mechanical properties such as the Cauchy pressure, $C_{11}-C_{44}$, and other quantities resulting from combinations of elastic constants. Poisson's ratio depends on both the crystallographic direction and the direction of deformation and is determined a combination of all three elastic constants $C_{11},C_{12}$ and $C_{44}$\,\cite{Baughman1998}. The differences in $C_{11}$ and $C_{12}$ lead to a maximal $\nu$ for Ag and Au of 0.81 and 0.882, respectively\,\cite{Baughman1998}. The minimum values of $\nu$ vary more drastically, reaching -0.091 for Ag and -0.029 for Au under uniaxial stress along the [110] direction. The negative values indicate auxetic behaviour in both Au and Ag, where lateral expansion occurs during longitudinal stretching along [110], as a consequence of many-body contributions to the binding energy\,\cite{Baughman1998}. Despite these variations, changes in $\nu$ have a limited effect on changes in $k$ related to the element. Moreover, the spring constant $k$ is highly sensitive to the local atomic configuration near the contact. Factors such as crystallographic orientation, defects and angular alignment of the contact with respect to a high symmetry direction, contribute to a distribution of different $k$ values, and thereby to spread measured values of $G_a$\,\cite{PhysRevLett.100.175502,PhysRevLett.87.026101,Rodrigues2000Au,PhysRevLett.100.175502}. Although there are slight differences in the spread of $G_a$ depending on the experimental run, as previously discussed in Ref.\,\cite{Calvo2018}, the observed difference in $G_{a,max}$ for Ag and Au is significantly larger than this spread. This difference directly reflects the distinct universal binding curves for the two elements, with Au exhibiting a stiffer bonding character compared to Ag. Thus, the magnetic energy exerts a stronger influence on $G_{a,max}$ in Ag compared to Au.

To estimate the magnetic contribution, we considered the susceptibility of bulk Au, where the dominant term is diamagnetic, with a value of $\chi_{Au}=-3.4\times 10^{-5}$ comparable to that of Ag\,\cite{PhysRevLett.108.047201,PhysRevLett.111.127202}). Taking into account the shape anisotropy of a cylinder, the anisotropic susceptibility is given by $\chi_{anis}=\chi_{\vert\vert}-\chi_{\perp} \approx 10^{-10}$\,\cite{PhysRevApplied.10.014030}. As described in the main text, the magnetic energy contribution can be estimated using the expression:  $E_{Magnetic}\approx \chi_{anis}V\frac{B^2}{\mu_0} sin(\theta)$. Assuming that the magnetically active volume can be approximated by a sphere of radius 100 nm, and that $\theta$ on the order of a few tens of degrees, we obtain $E_{mag}\approx 6 \times 10^{-7}$ eV. This value can be neglected as compared to other contributions in $E(z)$, indicating that the intrinsic magnetism of Au or Ag does not significantly affect the energies of the nanocontact. However, the repeated indentation process used to form atomic-scale contacts leads to the formation of nanometric wire-like structures of several tens of nanometers in length around the single-atom contact region \,\cite{PhysRevB.56.2154}. As discussed in the main text, macroscopic circulating currents within these anisotropic structures can result in much larger contributions to $E_{mag}$, consequently playing a significant role in modifying the energy landscape and contact formation under high magnetic fields.

\section*{Appendix B: Sticking and magnetism of O$_2$ at surfaces}

Studies of gold-oxygen chains have demonstrated that the strongest bonding arises from the formation of hybrid orbitals between Au and O$_2$\,\cite{PhysRevB.66.081405,PhysRevLett.96.026806}. In this context, the magnetic properties of O$_2$ are significantly influenced by the nature of the adsorption. While chemisorbed O$_2$ typically loses its magnetic moment, resulting in either a residual moment of 1 $\mu_B$ or complete quenching, physisorbed  O$_2$ retains a magnetic moment of 2 $\mu_B$\,\cite{doi:10.1021/acs.chemrev.7b00217,PhysRevLett.104.136101}. In the physisorbed configuration, the spin-up and spin-down states of the $\pi_p$ orbitals, which are the O$_2$ states closest to the Fermi level, remain spin-split, with an energy gap exceeding 1 eV between the spin-polarized bands\,\cite{min5040516}. In gold-oxygen structures, hybridized orbitals form, constituted primarily by Au $s$-electrons and minor contribution from O $p$-orbitals. These give rise to conduction bands with a small spin-splitted gap\,\cite{PhysRevB.66.081405}. However, there is no conclusive evidence for a significant influence of external magnetic fields on the conductance of the Au-O$_2$ chains. 

For gold, a magnetic moment of $1.3 \times 10^{-4} \mu_B$ per atom under an applied magnetic field of 10 T has been observed\,\cite{PhysRevLett.108.047201}. In addition, the paramagnetic susceptibility associated with the 5d orbital is on the order of $8.9 \times 10^{-6}$, and is known to exhibit a strong orbital contribution\,\cite{PhysRevLett.108.047201,Trudel2011}. These values are relatively small and therefore exclude the possibility of any significant intrinsic magnetic field effect on the electronic conduction through a single-atom Au point contact.

The force required to manipulate individual metal atoms on surfaces is highly dependent on the strength of their bonding to the substrate and can be on the order of pico Newtons (pN), even for strongly bonded metals. For example, a force of 17 pN has been reported for Co atoms on Cu surfaces\,\cite{doi:10.1126/science.1150288}. During the push-pull process associated with conductance versus distance measurements under applied magnetic fields, this force is sufficient to induce significant mobility in O$_2$ molecules. In the vicinity of the atomic contact, O$_2$ molecules benefit from the reduced coordination environment of the Au atoms, which enhances their ability to stick onto the contact region.

We estimate that the residual amount of O$_2$ in our experimental chamber to be of at least $10^{-7}$ mol, or approximately a partial pressure of O$_2$ of $10^{-9} mbar$. For context, previous studies of O$_2$ contacts at zero magnetic field were conducted with an O$_2$ concentration of ${10^{-5}}$\,mol\,\cite{PhysRevLett.96.026806,Thijssen_2008}. This leads to time scales for the formation of a monolayer of O$_2$ which can be very large, of many hours. We note that most of the O$_2$ is adsorbed on the cold surfaces around the sample. But the O$_2$ adsorbed close to the contact area should be affected by the repeated indentation procedure discussed above and in Refs.\,\cite{PhysRevLett.96.026806,Thijssen_2008}. This procedure leads to a considerable rearrangement of the contact area on a large spatial range, eventually favoring re-arrangements of adsorbed atoms. Furthermore, with the bias voltage applied here, large contacts formed a few times in between indentations may lead to currents in the mA range, giving very high current densities and a power dissipation of the order of fractions of a mW. This might lead to local heating effects eventually activating molecular motion. The combination of strong atomic rearrangement and heating seems helpful in enhancing molecular transport close to the contact area. Nevertheless, the exact mechanism by which additional molecules as O$_2$ come to the contact area requires further studies.

\section*{Appendix C: Magnetic anisotropy energy}
\begin{figure}[htbp]
    \centering
    \includegraphics[width = 0.95\columnwidth]{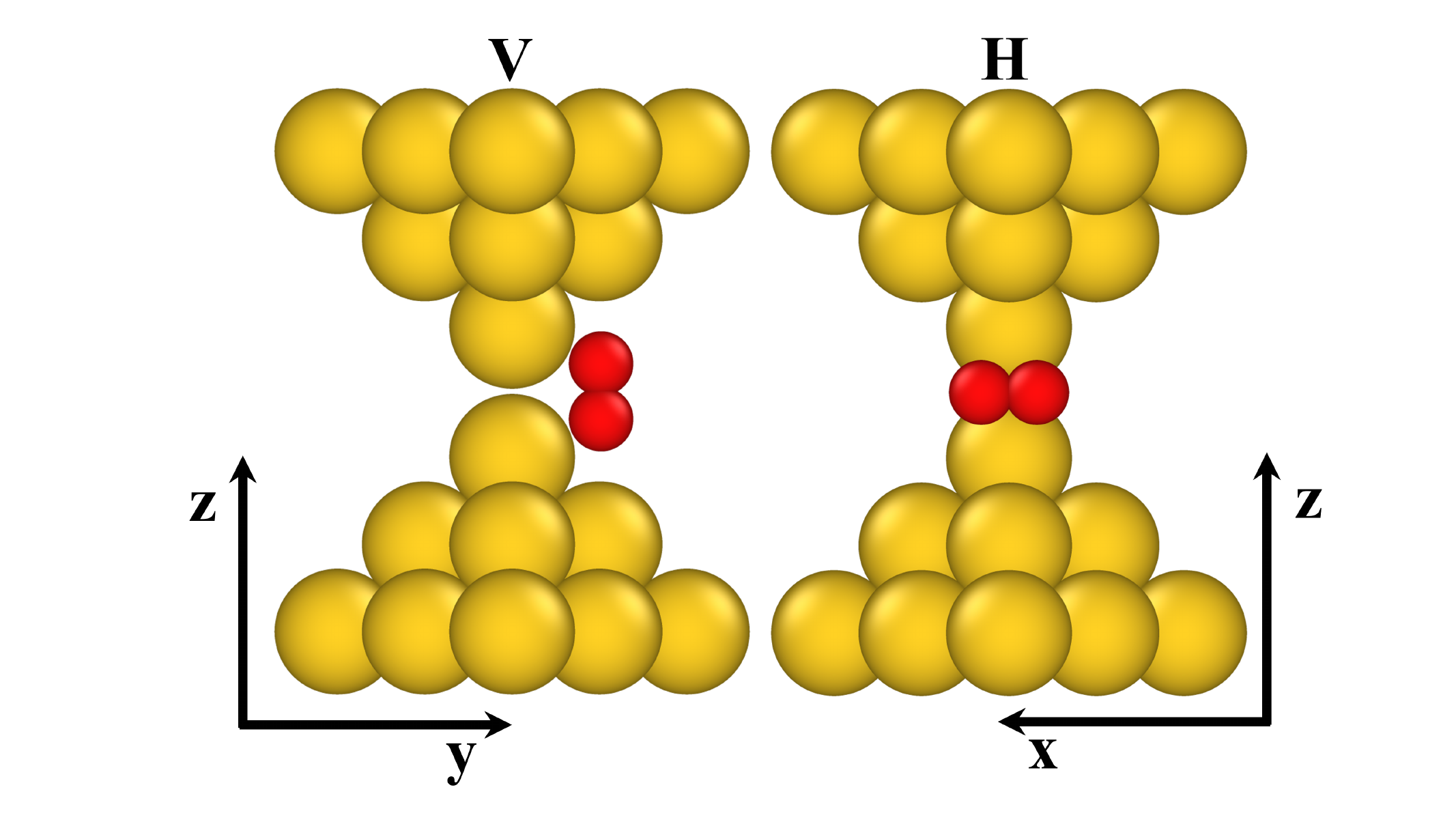}
    \caption{Schematic representation of the toy model used to estimate the MAE.}
    \label{FigSupp_ToyModel}
\end{figure}

The magnetic anisotropy energy (MAE) was estimated using a simplified model consisting of two gold pyramids oriented along the crystalline (001) direction, with atomic layers composed of 1-4-9 Au atoms. An O$_2$ molecule was positioned along the z-axis between the pyramids as illustrated in Fig.\,\ref{FigSupp_ToyModel}. Calculations were carried out using the new build-in ANT.Gaussian module for fully self-consistent Spin-Orbit Coupling (SOC). A modified basis set, Au\_pob\_TZPV\_rev2 \cite{Au_pob_TZVP_rev2, Stutt_pseudo}, was employed to consider SOC effects specifically in the central gold atoms of the toy model. Single-point energy calculation were performed with a convergence threshold of $3 \times 10^{-8}$ eV, with the spin axis fixed along the x,y and z directions, both in the presence and absence of SOC.

For model V (vertical arrangement of O$_2$, see Fig.\,\ref{FigSupp_ToyModel}), the energy difference between spin orientations in the absence of SOC is approximately $3 \times 10^{-7}$ eV, which is within the numerical accuracy of the calculations. It is negligible when compared to the difference in energy between the z-direction, which has the lowest energy  (estimated to be $6.5 \times 10^{-4}$ eV), and the x/y-direction, once the spin orbit coupling is considered. In model H (horizontal arrangement of O$_2$, also shown in Fig.\,\ref{FigSupp_ToyModel}), the energy difference without SOC is silightly larger, $6 \times 10^{-6}$ eV. When SOC is included, the x-direction becomes the most energetically favorable, with differences of $8 \times 10^{-5}$ eV relative to the y and z-directions. These results demonstrate the presence of an easy and hard axis in the system, induced on the molecule by the strong SOC of the Au atoms. Furthermore, these calculations also show that the magnitude and orientation of the MAE depend sensitively on the specific binding geometry of the O$_2$ molecule.

\section*{Appendix D: Conductance histograms}

\begin{figure*}[htbp]
    \centering
    \includegraphics[width = 0.8\textwidth]{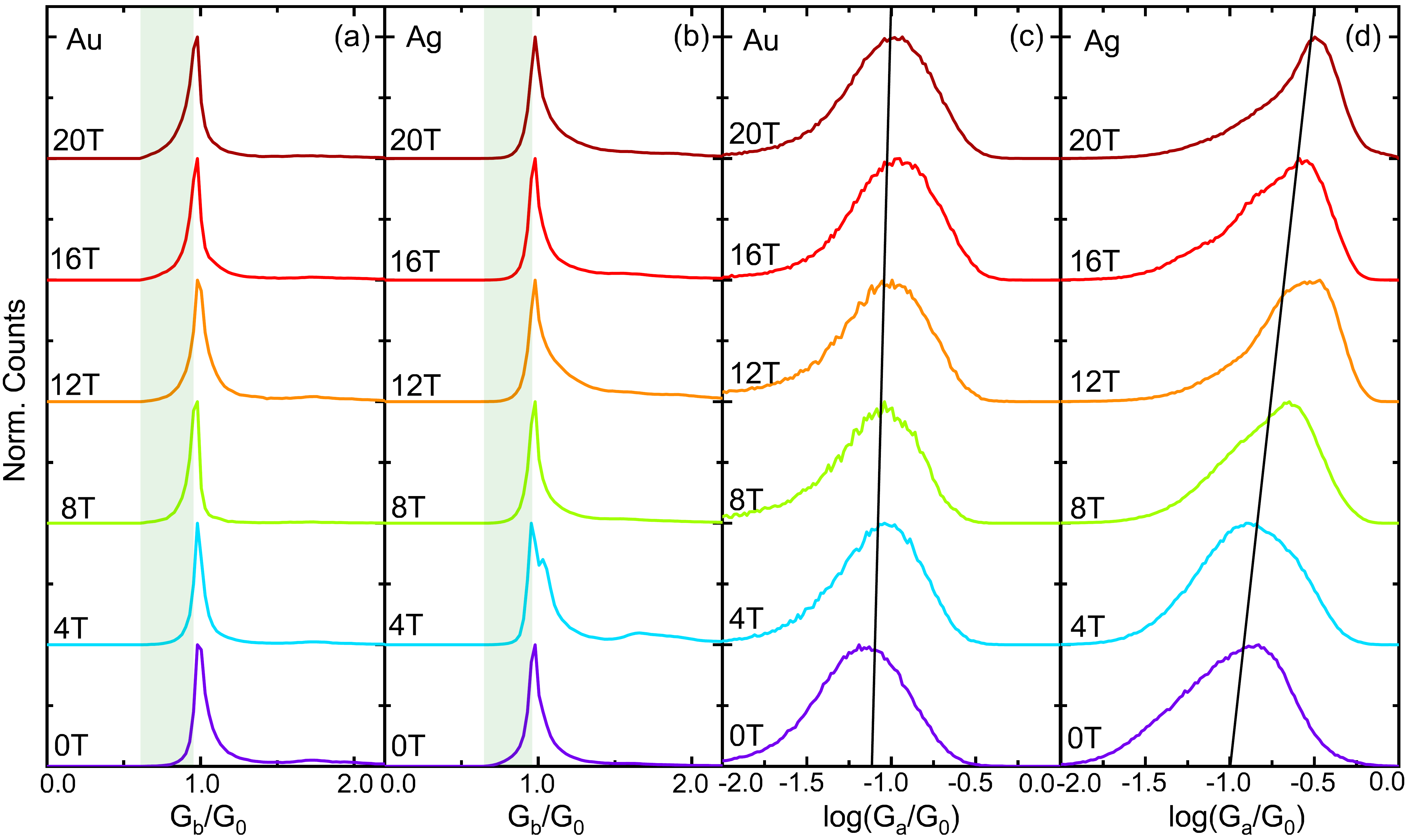}
    \caption{We show as colored lines the conductance histograms for fixed values of $G_a$ (a,b) and $G_b$ (c,d). Curves are shifted for clarity along the y-axis. The magnetic field is provided for each curve. In (a,c) we show Au contacts and in (b,d) Ag contacts. The histograms are normalized to one at the peak. In (a,b) we show the histogram as a function of $G_b$ for the value of $G_a$ at the peak of counts in Fig.\,\ref{FigureCurvesHisto}. In (c,d) we show the histogram as a function of $G_a$ for $G_b=1$. Note that the histograms here are just cuts for constant $G_a$ and $G_b$, as compared to Fig.\,\ref{FigureCurvesHisto}.  The number of counts is between 100 and 200 at the peak in Au (a,c) and between 100 and 500 at the peak in Ag (c,d).  The total number of conductance vs distance curves (each one with 2048 points) used to build the histograms are of about $10^5$ in Au and about 4 $\times$ $10^5$ in Ag at each magnetic field.}
    \label{FighistoGafixed}
\end{figure*}

We provide one-dimensional plots of the conductance histograms as curves in Fig.\,\ref{FighistoGafixed}. In Fig.\,\ref{FighistoGafixed}(a,b) we provide the conductance histograms for $G_a$ at the peak in the two-dimensional plot shown in Fig.\,\ref{FigureCurvesHisto}. We mark in green the area $G_b<G_0$, to highlight the increase in the number of data points showing $G_b<G_0$ at high magnetic fields. We see that the shape of the histogram is strongly asymmetric at zero field, as expected from the fact that the first contact has a well defined conductance at $G_0$, and becomes more symmetric at high fields, particularly for Au Fig.\,\ref{FighistoGafixed}(a). This is also seen in Fig.\,\ref{FigureHistoGaGb}(b) where we plot the ratio of contacts with conductance below $0.85 G_0$ as a function of the magnetic field. In Ag, the effect is much weaker.

Note that we observe a histogram with more features at 4 T in Ag. We have no clear explanation for this. It could be due to a peculiar situation close to the contact. Other curves do not show such features.

In Fig.\,\ref{FighistoGafixed}(c,d) we provide the conductance histograms for $G_b=1$, similar to the plot of Fig.\,\ref{FigureHistoGaGb}(a). We plot as black lines the lines shown in grey in Fig.\,\ref{FigureHistoGaGb}(a). We see that there is an ample set of occurrences of $G_a$. This is due to the varied atomic configurations surrounding the single atom point contact, each with different elastic properties, as discussed above. The histogram remains with the same shape in Au when increasing the magnetic field. For Ag, we observe that the maximum shifts clearly to higher values with the magnetic field. We also observe that the histogram at zero field is slightly skewed, and that the skewness becomes more important at high magnetic fields. We have no clear explanation of this behavior. Possibly, it can be related to the influence of magnetic effects in the interaction potential (Fig.\,\ref{FigSupplModel}).

\bibliographystyle{apsrev4-1-titles}

%

\end{document}